\pdfoutput=1 
\documentclass[fleqn]{mnras}

\usepackage{newtxtext,newtxmath}

\usepackage[T1]{fontenc}
\usepackage{ae,aecompl}

\usepackage{graphicx}	
\usepackage{amsmath}	
\usepackage{amssymb}	

\hypersetup{pdfauthor={C.Ceccarelli},
               pdftitle={The evolution of grain mantles and silicate dust growth at high redshift},
               pdfkeywords={galaxies: high-redshift -- (ISM:) ices -- (ISM:) dust},
               bookmarksnumbered=true}

\title[Cosmic ices]{The evolution of grain mantles and silicate dust growth at high redshift}

\author[C. Ceccarelli et al.]{
Cecilia Ceccarelli$^{1,2}$,\thanks{E-mail: cecilia.ceccarelli@univ-grenoble-alpes.fr}
Serena Viti$^{3}$,
Nadia Balucani$^{4,1,2}$,
Vianney Taquet$^2$
\\
$^1$Univ. Grenoble Alpes, CNRS, IPAG, F-38000 Grenoble, France\\
$^2$INAF-Osservatorio Astrofisico di Arcetri, Largo E. Fermi 5, I-50125, Florence, Italy\\
$^3$Department of Physics and Astronomy, University College London, Gower St., London, WC1E 6BT, UK\\
$^4$Dipartimento di Chimica, Biologia e Biotecnologie, Via Elce di Sotto, 8, Perugia, 06123, Italy
}

\date{Accepted XXX. Received YYY; in original form ZZZ}

\pubyear{2017}

\begin{document}
\label{firstpage}
\pagerange{\pageref{firstpage}--\pageref{lastpage}}
\maketitle

\begin{abstract} 
  In dense molecular clouds, interstellar grains are
  covered by mantles of iced molecules. The
  formation of the grain mantles has two important consequences: it
  removes species from the gas phase and promotes the synthesis of
  new molecules on the grain surfaces. The
  composition of the mantle
  is a strong function of the environment which the cloud belongs
  to.  Therefore, clouds in high-zeta galaxies, where conditions
  -like temperature, metallicity and cosmic rays flux-
  are different from those in the Milky Way, will have different grain
  mantles. In the last
  years, several authors have suggested that silicate grains might
  grow by accretion of silicon bearing species on smaller seeds. This
  would occur simultaneously to the formation of the iced mantles and
  be greatly affected by its composition as a function of time. In
  this work, we present a numerical study of the grain mantle
  formation in high-zeta galaxies and we quantitatively address the
  possibility of silicate growth. We find that the mantle thickness
  decreases with increasing redshift, from about 120 to 20 layers for
  $z$ varying from 0 to 8. Furthermore, the mantle composition is also
  a strong function of the cloud redshift, with the relative
  importance of CO, CO$_2$, ammonia, methane and methanol highly
  varying with $z$. Finally, being Si-bearing species always a very
  minor component of the mantle, the formation of silicates in
  molecular clouds is practically impossible.
\end{abstract}





\section{Introduction}\label{sec:introduction}

As it is well known, the interstellar medium (ISM) is made up of gas
and dust, the latter composed of silicates and carbonaceous material
(Hoyle \& Wickramasinghe 1969; Jones et al. 2013, 2017). In recent
years, a heated debate has arisen on the quantity of silicates in the
high redshift ISM and on its formation (Draine 2009; Michalowski et
al. 2010; Mancini et al. 2015; Zhuvkovska et al. 2016; Ginolfi et
al. 2018). In the local ISM, the major sources of silicates are
believed to be the Asymptotic Giant Branch (AGBs) stars and, at to
lesser extent, supernovae (SNe), although large uncertainties exist
about how much dust is formed in both cases (e.g. Nanni et
al. 2013; Ventura et al. 2014; Sugerman et
al. 2006; Barlow et al. 2010; De Looze et al. 2017).
At high redshifts, the situation is less clear, as low-mass stars,
in principle, do not have time to evolve into AGBs (but see Valiante et
al. 2009, 2011). Therefore, some authors have argued that the major
source of high-$z$ silicates are SNe (Kozasa et al. 1991; Todini \&
Ferrara 2001; Schneider et al. 2004; Bianchi \& Schneider 2007; Dwek
et al. 2007).
Alternatively, others claim that silicates might
grow, meaning that they form, and not only coagulate from smaller grains, in the
cold and dense molecular clouds, via the freeze-out of atomic silicon
or SiO molecules (e.g. Valiante et al. 2011; Asano et al. 2013;
Hirashita \& Voshchinnikov 2014; Mancini et al. 2015; Ginolfi et
al. 2018). All these studies implicitly assume that gaseous Si and SiO would
accrete into small silicate seeds which can then grow. However,
there are two difficulties with this assumption.

First, the formation of silicates is unlikely to occur in cold
environments (e.g. Goumans \& Bromley 2012). Several laboratory
experiments show that SiO easily dimerises, even at low temperatures,
if a seed of TiO$_2$ or silicates are already present. SiO molecules
can also form clusters that can eventually condense as solid silicon
oxide (see e.g. Krasnokutski et al. 2014 and references
there). However, neither SiO dimers or clusters form ``real''
silicates. Besides, SiO dimers or clusters do not have the correct
silicon/oxygen ratio to become silica (based on SiO$_2$ crystals) or
silicate (SiO$_4$ crystals).
The second difficulty in the ``Si, SiO freeze-out = silicate growth''
scheme is the following. Even if, via some unknown process, the
formation of silicate occurs in cold environments, it is unlikely
that the accreted Si or SiO will meet on the surface and 
form the necessary dimers or clusters. In a recent study,
Ferrara et al. (2016) analytically argued that, at redshift of $\sim$
6, SiO molecules will be prevented from getting in contact with the
silicate surface and from forming silicate-like bonds because of the
formation of icy mantles, but no specific modelling was carried out.

While we cannot say more than what is already known about the silicate
formation from a chemical point of view (which is summarised above),
we can compute the probability that two or more Si and SiO species
meet on the grain surface to form dimers or clusters. Whether they
would then build up a silicate-like bond remains speculative at this
point. Nonetheless, it is first important to assess whether there is
a non-zero the probability to have dimers and clusters in the first
place.  In practice, the formation of Si or SiO dimers and clusters
might only occur in the cold molecular clouds (see Ferrara et
al. 2016), when the grain mantles form. The question is then what is
the abundance of Si and SiO on each layer of the mantle with respect
to other mantle species. If, for example, in one layer we have 1/100
SiO molecules with respect to the other frozen molecules, the
probability that two SiO are close enough to form a dimer is
$10^{-2}$ and even less for a SiO cluster.  {\it Therefore, the possible
  grain growth is closely connected with the formation of the grain
  iced mantles process.}

ISO, AKARI, and Spitzer satellites have provided us with a plethora of
NIR observations on the composition of dust ices towards Galactic star
forming regions and a few external nearby galaxies. These observations
show that the interstellar grain mantles are prevalently made of water
ice, with a smaller fraction of CO, CO$_2$, CH$_4$, NH$_3$, and
CH$_3$OH, at a level of 5-30\% each (Boogert et al. 2015 and
references therein). The current explanation for the relatively high
abundance of these iced species (when compared to their gas-phase
abundances) is that hydrogenation and, possibly, oxidation of atoms
(C, O and N) and simple molecules (CO) takes place on the grain
surfaces. For example, water is believed to be the result of
hydrogenation of the oxygen atoms landing on the grain surfaces
(Dulieu et al. 2010; Lamberts et al. 2014), while formaldehyde and
methanol are the hydrogenated products of CO (e.g. Watanabe \& Kouchi
2002; Rimola et al. 2014; Song \& Kastner 2017).

In addition to having a crucial role in the possible grain growth, the
icy mantles formation heavily influences the chemical and physical
evolution of the gas and dust in the clouds, because reactants and
coolants are removed from and then re-injected into the gas
phase. In turn, the icy mantles formation depends on a complicated
time-dependent, nonlinear chemistry highly affected by the physical
environment.  While several theoretical studies have concentrated on
exploring how the gas chemical composition may vary with the different
physical conditions that characterise the ISM of external galaxies,
such as metallicity, gas density, cosmic ray ionisation rate, and
far-UV radiation intensity (e.g. Bayet et al. 2009, 2011; Meijerink et
al. 2009, Bisbas et al. 2015; Acharyya \& Herbst 2016), no theoretical
study on how ices change with such parameters has been carried out so
far.

With this article, we aim to fill up this gap.  Its focus is the study
of the formation of the grain iced mantles at high redshift $z$,
with the twofold scope to understand how they change with $z$ and to
quantify the probability that SiO dimers or clusters form and,
eventually, grow into silicates.
To this end, we revise the chemical reactions involving Si-bearing
species that can occur on the grain surfaces and model the evolution
of ices as a function of several physical parameters that, as a
function of redshift, are likely to differ from galactic values
(Sec. \ref{sec:chemical-modelling}). We then explore the abundances of
the main ice components as well as Si-bearing solid species in order to
quantify whether grain growth at high redshift can indeed be efficient
enough to account for the potential missing dust mass
(Sec. \ref{sec:results}). 
Finally, in Sec. \ref{sec:conclusions}, we comment on the implications
of our study.

\section{Chemical Modelling}\label{sec:chemical-modelling}

The code that we use in this study is based on the GRAINOBLE model
(Taquet et al. 2012, 2013). It is a time-dependent grain-gas chemistry
code that also allows us to follow the structure of the layered grain
mantles as a function of time.  The code used here adopts the
formalism by Hasegawa \& Herbst (1993) for computing the surface
reactions, keeping the layered structure composition (Taquet et
al. 2014).  Briefly, the code computes the abundance of grain surface
and gas phase species as a function of time. In the following, we give
a short summary of the processes included in the code, the updated
chemical network that we use for this study, and describe the
parameters of the modelled cloud.

\subsection{Processes}\label{sec:processes}

{\it Accretion:} Species can accrete from the gas to the
grain surfaces at a rate $k_{accr}$, following the standard formalism:
\begin{equation}
  \label{eq:1}
k_{accr} = S_x \pi a^2 n_{grain} v_x         
\end{equation}
where $S_x$ is the sticking probability of the accreting species $x$,
$a$ the (average) grain radius, $n_{grain}$ the grain number density,
$v_x$ the velocity of the gaseous species $x$ with mass $m_x$ and
kinetic temperature $T_{gas}$, and $v_x = (2 k_B T_{gas}/m_x)^{1/2}$
where $k_B$ is the Boltzmann constant). Following the experimental
study by He, Acharyya and Vidali (2016a), we assumed that the sticking
coefficient $S_x$ is equal to unity for all species except atomic
hydrogen, for which we followed the formalism in Tielens (2005). 

  \begin{table}
    \begin{tabular}{llclc}
      \hline
      \hline
      No. & Reactants &                     & Products & E$_{act}$ [K]\\
      \hline
      1    & Si + O  & $\rightarrow$ &  SiO  & 0 \\
      2    & SiO + O   & $\rightarrow$ & SiO$_2$ & 1000\\
      3    & SiO + OH & $\rightarrow$ & SiO$_2$ + H & 0 \\ 
      \hline
      Species & E$_b$ [K] & &  & \\
      Si & 2700  & &  & \\
      SiO & 3500  & &  & \\
      SiO$_2$& 4300  & &  & \\
      \hline
      \hline
    \end{tabular}
    \caption{{\it Top half:} List of reactions occurring on the grain surfaces
      involving the oxidation of SiO. Last column reports the value of
      the energy barrier in K, as computed by Martin et
      al. (2009). {\it Lower half:} Binding energies of the Si-bearing
      species considered in this study, as measured by He at al. (2016b).}\label{tab:SiO2}
  \end{table}

\vspace{0.4cm}
\noindent {\it Diffusion:} Once on the grain surface, the species can
diffuse via the standard thermal hopping process. The diffusion energy
is assumed to be 0.5 times the binding energy E$_b$ of the species. The
binding energies are listed in Taquet et al. (2012, 2014) and are
updated according to He, Acharyya and Vidali (2016b). For Si, SiO
and SiO$_2$, the main species of interest in this work, we give
the adopted value of their binding energies in Table \ref{tab:SiO2}.

\vspace{0.4cm}
\noindent {\it Surface reactions:} Species on the grain surfaces can
meet, react and form other species, following the Langmuir-Hinshelwood
mechanism. The reaction rate is given by the product of the diffusion
rate (see above) and the probability of the reaction. For
radical-radical reactions the latter is equal to unity, otherwise it
is a function of the reaction energy barrier E$_{act}$ and it is calculated
following the Eckart model (see Taquet et al. 2013 for more
details). Note that H-tunneling can make the reaction efficient,
despite the presence of a barrier (e.g. Rimola et al. 2014). GRAINOBLE
considers that the reaction can only occur on the species within the
last two mantle layers (i.e. the surface), namely the bulk of the
mantle is assumed chemically inert.

\vspace{0.4cm}
\noindent {\it Desorption:} Species on the grain surfaces can be
injected into the gas phase via thermal desorption, cosmic-ray induced
desorption and chemical desorption, in each case following the
relevant standard formalism (e.g. Taquet et al. 2012). The first two
desorption mechanisms depend on the species binding energy (see
above), while for the latter we assumed that 1\% of species formed on
the grain surfaces are injected back into the gas phase (Garrod et
al. 2007). \\

\vspace{0.4cm}
\noindent {\it Gas-phase reactions:} Species in the gas undergo
two-body reactions as the three body reactions are extremely rare at
the considered densities.

\subsection{Chemical networks}\label{sec:chemical-networks}

For the reactions occurring in the gas phase, we used the KIDA 2014
network (Wakelam et al. 2015; {\it http://kida.obs.u-bordeaux1.fr}),
updated with the reactions described in Loison et al. (2014), Balucani
et al. (2015) and Skouteris et al. (2017). 

The reactions on the grain surfaces leading to formaldehyde, methanol
and water are described in Taquet et al. (2013) and Rimola et al.
(2014). In addition, we inserted the hydrogenation of C, N and Si
atoms, which has no energy barrier and leads respectively to methane (CH$_4$),
ammonia (NH$_3$) and silane (SiH$_4$; and the respective partially hydrogenated
species)\footnote{In absence of
  specific quantum-chemistry
  computations or laboratory experiments, we assumed that the back
  reactions, like for example CH$_2$ + H $\rightarrow$ CH + H$_2$,
  do not take place, so that the final hydrogenated products might be
  slightly overestimated.}.  

Finally, in the grain surface reaction network, we included the
oxidation of Si and SiO, which leads to SiO and SiO$_2$,
respectively. The first reaction, Si + O $\rightarrow$ SiO, is
barrierless. For the oxidation of SiO we considered the reactions with
O and OH. The first one has a large barrier, whereas the second
reaction is barrierless (contrarily to the CO + OH reaction) and has a
rate of $6\times 10^{-12}$ cm$^{-3}$s$^{-1}$ (Martin et al. 2009).
Table \ref{tab:SiO2} summarises the SiO oxidation reactions and the
binding energies E$_b$ of the Si-bearing species considered in this
study.

\subsection{Cloud temperature, elemental abundances and other
  parameters}\label{sec:elem-abund-other} 

We model a molecular cloud in a galaxy at a given redshift $z$. We
assume that the cloud is completely shielded from the UV photons of
the interstellar radiation field and that the H number density of the
molecular cloud, $n_H$, is constant and equal to $2\times10^4$
cm$^{-3}$. In the simulations, we assume an average grain radius of
0.1 $\mu$m\footnote{Note that very small grain, $\leq0.02$ $\mu$m,
    are stochastically photon heated and the mantles do not survive
    (e.g. Hollenbach et al. 2009).}, typical of the galactic ISM grains
  (e.g. Jones et al. 2013). 

  The dust and gas are assumed to
  be thermally coupled, which, at the considered density and
  conditions, is approximately correct (e.g. Goldsmith 2001; Gong,
  Ostriker \& Wolfire 2017). The temperature is constant with time and
  depends on the redshift of the molecular cloud. Specifically, we
  assumed that the cloud temperature is $\sim$10 K larger than the
  Cosmic Background Radiation (CMB) temperature at a given redshift to
  account for the additional dust heating from the interstellar
  photons field (Table \ref{tab:parameters}). We note that $\sim$10-12 K are the average gas and dust temperatures in galactic molecular clouds. 
 
The abundances of the elements in the gas phase are assumed to be
the solar ones with the following modifications: (1) oxygen and
carbon are assumed to be half on the refractory grains and half in
the gas; (2) 90\% of chlorine is assumed to be depleted while 10\%
is in the gas phase (which may be a pessimistic assumption, as Cl is
not in rocky materials; see e.g. the discussion in Codella et
al. 2012); (3) F, P, Fe, Mg, Al, Ca and Na are largely depleted in the
refractory grains and only 1\% is in the gas phase; (4) Si is a
parameter of this study and we varied it from 90\% to 1\% of it
being in the gas phase, in order to verify whether silicates can
grow in molecular clouds. Table \ref{tab:elements} summarises the
adopted elemental abundances.

\begin{table}
  \begin{tabular}{lccccc}
    \hline
      \hline
    Element & unit & \multicolumn{4}{c}{Abundance wrt H}\\
                 &  & Ref. & $met$0.1 & $met$0.01 & $Cosmo$ \\ 
    \hline
    H   & 1 & 1 & 1 & 1 & 1 \\
    He & 1 & 0.85 & 0.85 & 0.85 & 0.92  \\
    O   & [$\times 10^{-4}$] & 2.5 & 0.25 & 0.025 & 1.8   \\
    C   &  [$\times 10^{-4}$] & 1.3 & 0.13 & 0.013 & 0.55  \\
    N   &  [$\times 10^{-5}$] & 6.8 & 0.68 & 0.068  & 2$\times 10^{-6}$  \\
    Cl   &  [$\times 10^{-7}$] & 1.5 & 0.15 & 0.015 & 0.02 \\
    F   &  [$\times 10^{-7}$] & 2.0 & 0.20 & 0.020 & 5$\times 10^{-7}$  \\
    Fe & [$\times 10^{-7}$] & 3.5 &0.35 & 0.035 & 2.5  \\
    Mg  & [$\times 10^{-7}$] & 4.0 &0.40 & 0.040 &  6.5 \\
    Na   &  [$\times 10^{-8}$] & 1.2 &0.12 & 0.012 & 0.97 \\
    P   &  [$\times 10^{-7}$] & 2.6 & 0.26 & 0.026 & 0.065 \\
    S    & [$\times 10^{-5}$] & 1.3 & 0.13 & 0.013 & 0.03  \\
    Si  & [$\times 10^{-7}$] & 3.2$^*$ &0.32& 0.032 & 32   \\
      \hline
    \hline
  \end{tabular}
  \caption{List of the gaseous elemental abundances, with respect to H
    nuclei, used in the
    simulations. Column Ref. lists the abundances of our reference
    model. Columns $met0.1$ and  $met0.01$ refer to a metallicity
    0.1 and 0.01 with respect to the solar one, namely all species are
    scaled by that factor (see text for details). Column
    $Cosmo$ lists the abundances as derived by cosmological
    simulations (see text ). Note: $^*$The elemental Si abundance is
    a parameter varied in the simulations, as reported in Table
    \ref{tab:parameters}. }
  \label{tab:elements}
\end{table}

The final mantle chemical composition depends also on two additional
parameters: the cloud metallicity, $met$, and the cosmic-ray (CR)
ionisation rate, $\zeta_{CR}$. 
Thus, we explored a range of metallicity from 0.01 to 1 the solar one
by reducing the elemental abundances by the same factor. We also
consider a case where the elemental abundances are taken from
simulations of the ISM enrichment from Pop III stars (see Table
\ref{tab:elements} and Section \ref{sec:effect-met}).
Finally, CR ionisation rate is governed by the rate of Supernovae
explosions (e.g. Morlino G. 2017), which might be a function of the
redshift. Therefore, we run models covering the range
$1\times 10^{-17}$ s$^{-1}$, equivalent to a low value in the Milky
Way (Padovani et al. 2009), to $1\times 10^{-14}$ s$^{-1}$,
corresponding to a cloud irradiated by a 1000 times more intense CR
flux (Vaupr\'e et al. 2014), which might simulate the conditions in
early galaxies.

Table \ref{tab:parameters} summarises the parameters used in our
simulations and the adopted values.

\begin{table}
  \begin{tabular}{lcc}
    \hline
    \hline
    Parameter & Range of values & Model number\\
    \hline
    $z$                   &  0,  3, {\bf 6}, 8  & \\
    Temperature [K] & 10, 20, {\bf 30}, 35 & 1--4 \\
    $\zeta_{CR}$ [$\times 10^{-17}$ s$^{-1}$] & {\bf 1}, 10, 100, 1000  & 5--8\\
    $met$  & $Cosmo$, 0.01, 0.1, {\bf 1} & 9--12\\
    $Si_{gas}/Si_{sol}$ & {\bf 0.01} 0.1, 0.3, 0.9 & 13--16\\ 
    \hline
    \hline
  \end{tabular}
  \caption{Parameters varied in the simulations. We considered
    values aimed to simulate the conditions of clouds at redshift $z$
    0, 3, 6 and 8, quoted in the first row, which correspond to the (dust and gas)
    temperatures of the second row. The temperature was computed by 
    adding $\sim$10 K to the CMB one at a given redshift $z$ and,
    specifically, varies from 10 to 35 K (see text). 
    The CR ionisation rate $\zeta_{CR}$ varies  
    from the Milky Way low value, $1\times 10^{-17}$ s$^{-1}$, to
    1000 times larger, likely appropriate for early galaxies. 
    The metallicity $met$ varies from 0.01 to 1 of
    the elemental abundances listed in Table
    \ref{tab:elements}, to simulate different enrichment. In
    addition, we run a case with elemental abundances predicted by
    cosmological simulations, model $Cosmo$ (see text).
    Finally, the elemental silicon abundance in the gas phase, $Si_{gas}$,
    varies from 90\% (namely $2.9\times10^{-5}$) to
    1\% ($3.2\times10^{-7}$) of the solar value, to explore whether
    silicates can grow in molecular clouds. The values in 
    boldface font refer to our reference model (see text). The last
    column reports the number of each model, used in Table
    \ref{tab:tablone}. }
  \label{tab:parameters}
\end{table}

\subsection{Chemical evolution}\label{sec:chemical-evolution}

We started with a partially atomic cloud, meaning that all elements
are in the atomic form (neutral or ionised depending on their
ionisation potential) except hydrogen, which is assumed to be in
molecular form. This mimics the pseudo-evolution of a cloud from
neutral to molecular. The chemical composition of the cloud is left to
evolve with time for 10 millions years. Note that GRAINOBLE follows
the chemical evolution of each grain mantle layer, so that it provides
the composition of each layer as the mantle grows. In general, the
mantle is formed of about 100 layers, the exact number depending on
the temperature, the CR ionisation rate, and the elemental abundances
(especially O, C and N) assumed in the model (see Results).

\section{Results}\label{sec:results}

In this section we shall investigate the composition of each layer of
the icy mantles, for a comprehensive parameter space. To this scope,
all the figures shall plot the fractional abundance (with respect to
the total number of H nuclei) of selected frozen species as a function
of time. Plots are reported for three sets of frozen species. The
first set is formed by the major constituents of the grain mantles:
H$_2$O, CO, CO$_2$, CH$_4$ and NH$_3$. The second set plots minor
frozen species: CH$_4$, CH$_3$, H$_2$CO and OH. Finally, the third set
is focused on the Si-bearing frozen species, for assessing the
possibility of silicate growth: SiO$_2$, SiO, SiH$_4$ and Si.

In order to evaluate whether grain growth can occur, it is important
to show the ice evolution as a function of the building up of the
layers. Note that for Si-bearing species to be able to form bonds on
the core grain and, therefore, possibly form silicate-like bonds
(see Introduction), they would have to be abundant in the first few
layers. One can also think of a situation where small clusters of Si
and SiO or SiO dimers can form in deeper layers and then they
remain clustered when the mantle sublimate and, therefore, constitute
small silicate seeds. In this case, in order to grow silicates, frozen Si and
SiO have to be abundant and close to each other.

In the following, we will discuss separately the effects due to the
different parameters.

\subsection{The effect of the CMB temperature}\label{sec:effect-cmb}

In this section, we investigate the effect of increasing the CMB
temperature on the icy mantle composition. We assumed that the cloud
temperature is about 10 K larger than the CMB blackbody spectrum
temperature, namely $T_{CMB} = T_0(1 +z)$, with $T_0$ = 2.725 K. We,
therefore, considered 10, 20, 30 and 35 K to simulate the clouds at
$z$ equal to 0, 3, 6 and 8 (Table \ref{tab:parameters}).

\begin{figure}
\includegraphics[width=8cm]{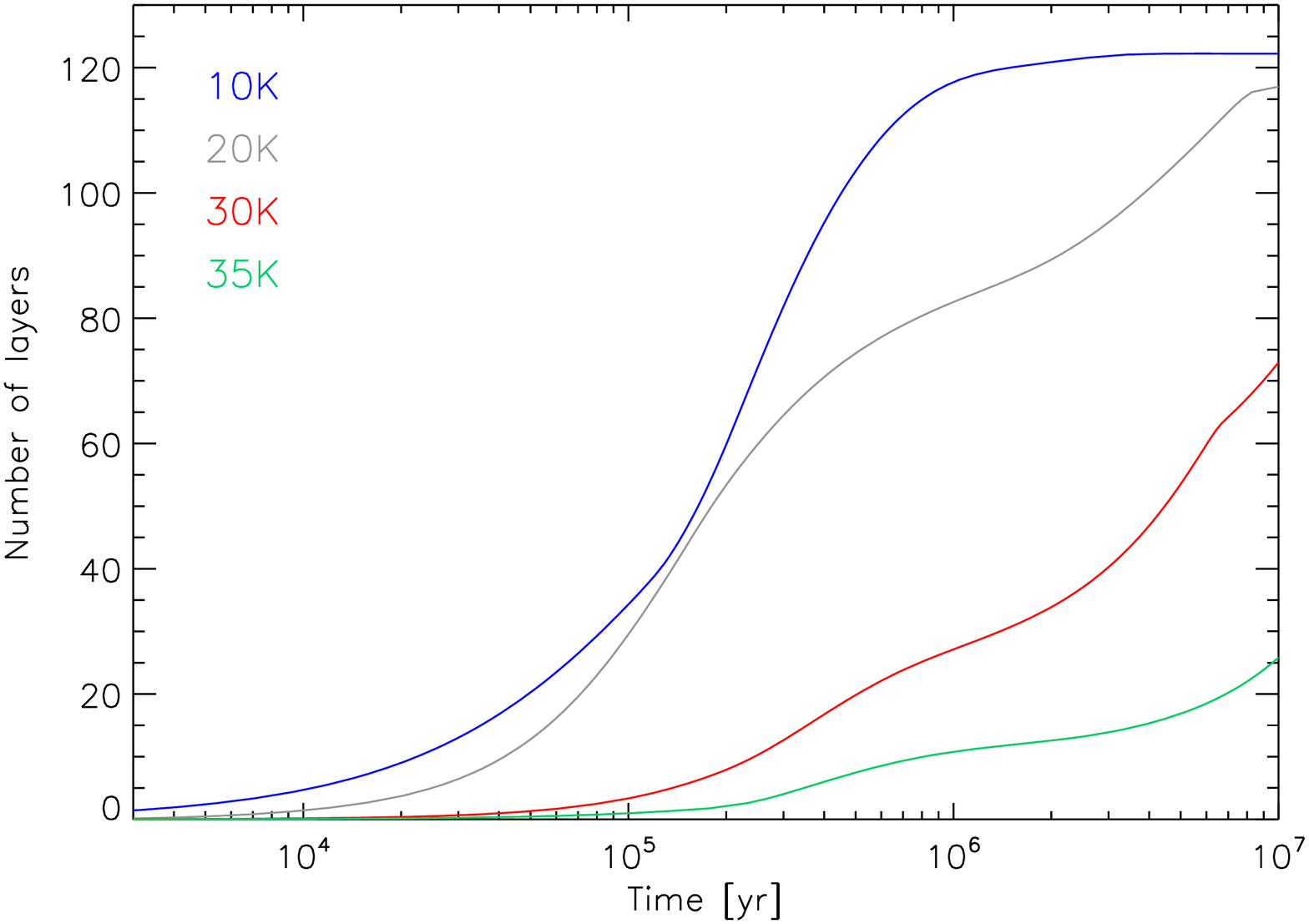}
\caption{Number of layers of the grain mantle as a function of time
  for the four models with different cloud temperature: 10 K (blue),
  20 K (grey), 30 K (red: this is the reference model) and 35 K
  (green). }
\label{fig:CMB-Nlayers-time}
\end{figure}

Figure \ref{fig:CMB-Nlayers-time} shows the number of layers composing
the grain mantle as a function of time and for the different
temperatures. Already from this figure it is clear that the
temperature, namely the redshift of the cloud, has a great influence
on the grain mantle growth: the larger the temperature the smaller the
number of mantle layers.
\begin{figure*}
\includegraphics[width=14cm]{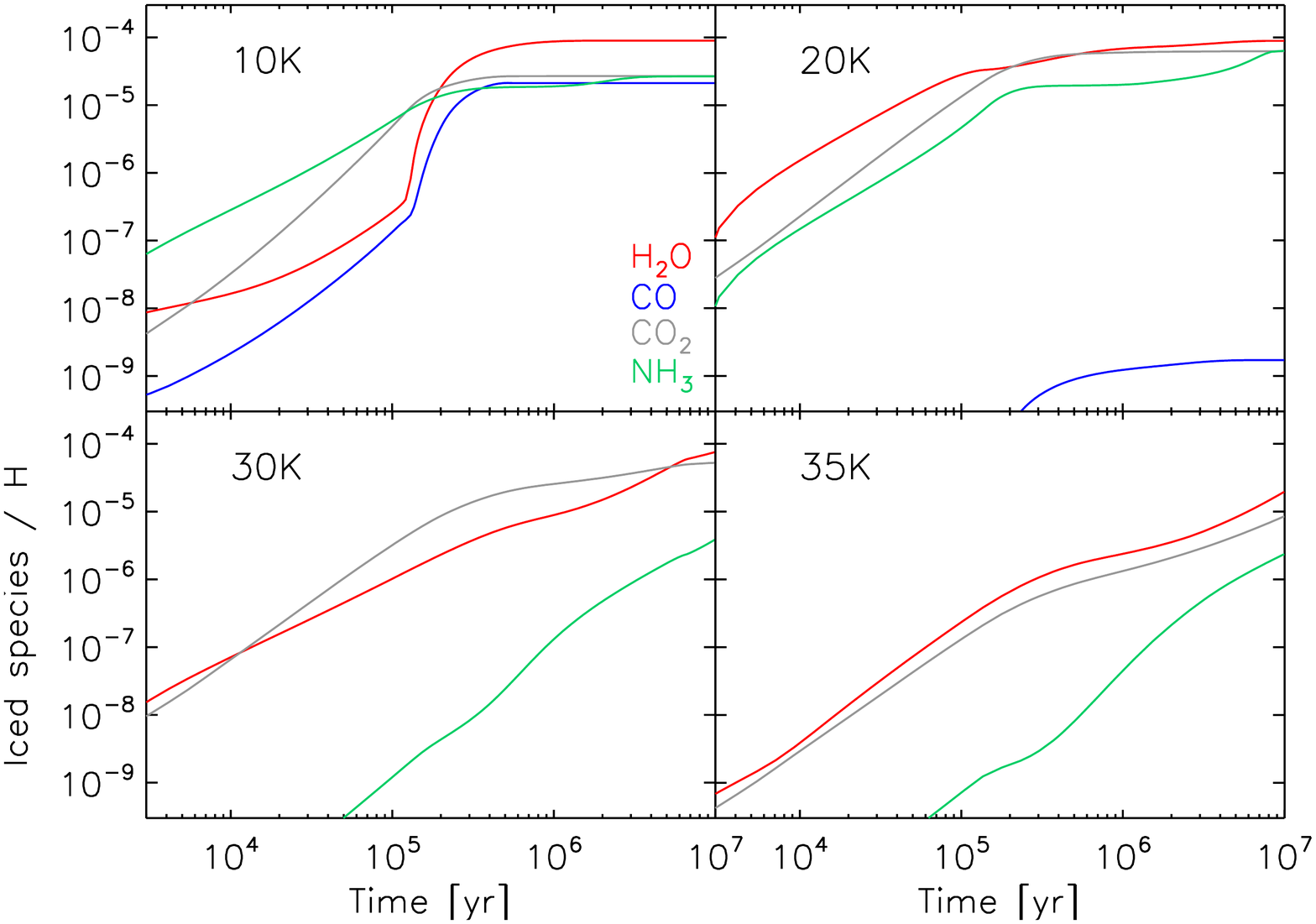}
\caption{Grain mantle chemical structure.
  Cumulative fractional abundances with respect to the total
  number of hydrogen nuclei of iced water (red), CO (blue), CO$_2$
  (grey) and ammonia (green) as a function of time. The cloud
  temperature is 10 K ({\it upper left panel}), 20 K ({\it upper right
    panel}), 30 K (({\it bottom left panel}) and 35 K ({\it bottom
    right panel}). }
\label{fig:CMB-Mantle1}
\end{figure*}

\begin{figure*}
\includegraphics[width=14cm]{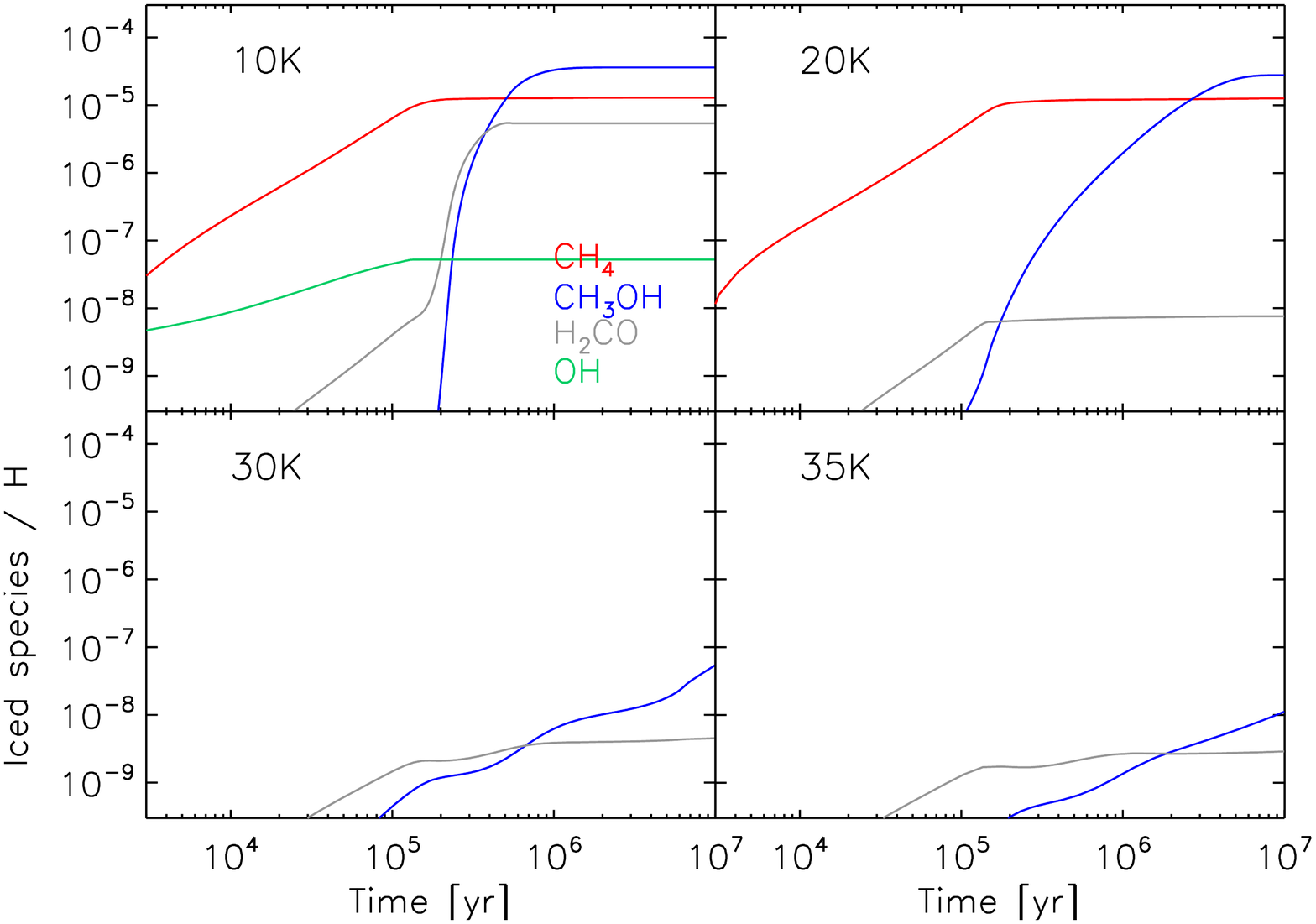}
\caption{As Fig. \ref{fig:CMB-Mantle1} for iced methane (red), methanol
  (blue), formaldehyde (grey) and OH (green).}
\label{fig:CMB-Mantle2}
\end{figure*}

\begin{figure*}
\includegraphics[width=14cm]{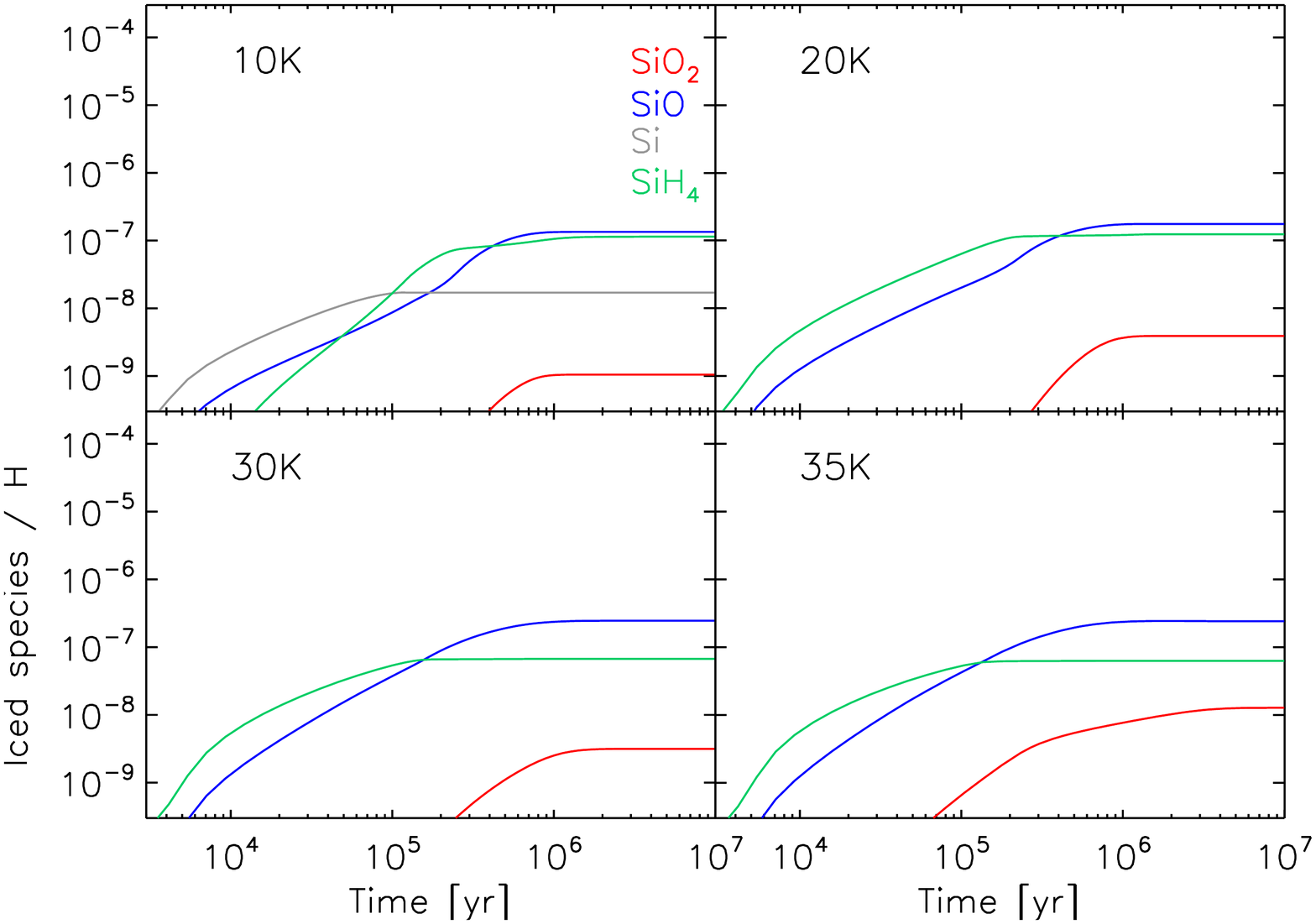}
\caption{As Fig. \ref{fig:CMB-Mantle1} for iced SiO$_2$ (red), SiO
  (blue), atomic Si (grey) and silane (green).}
\label{fig:CMB-Mantle3}
\end{figure*}

In order to understand how the redshift specifically affects the
mantle composition, we first start with a quick analysis of the grain
mantle composition of a ``standard'' Milky Way cloud, namely a cloud
with a temperature of 10 K (Figs. \ref{fig:CMB-Mantle1} to
\ref{fig:CMB-Mantle3}).
The mantle is composed by about 120 layers which grow in about
2--10$\times10^5$ years (Figure \ref{fig:CMB-Nlayers-time}). The first
layers are mostly composed by the hydrogenated forms of C and N,
namely methane and ammonia. This is because our model starts with an
atomic cloud, and when the C and N atoms land on the grain they
undergo hydrogenation, which is a fast process since it does not have
energy barriers. After $\sim 10^5$ yr, the gaseous abundance of both C
and N drops, as CO and N$_2$ form.
On the other hand, the hydrogenation of oxygen takes more time than
that of C and N because of the larger binding energy of O ($\sim1400$
K against $\sim800$ K of C and N: e.g. Bergeron et al. 2008).
In addition, at early times ($\leq 10^5$ yr), atomic oxygen is more
abundant than atomic hydrogen while iced CO starts to increase, so
that the formation of iced CO$_2$ occurs before that of water (which
is not a simple barrierless process).  The first 35 mantle layers
are, therefore, mostly constituted of ammonia, methane and CO$_2$.
Between about 1 and 3 $\times 10^5$ yr the abundances of atomic oxygen
and hydrogen become comparable, and water formation takes over.
The intermediate mantle layers are, therefore, dominated by water
molecules. Finally, at $\geq 4\times 10^5$ yr, the gaseous CO
freeze-out into the mantles and methanol is efficiently formed by its
hydrogenation. The last 35 or so layers are a mix of water, CO and
methanol. 
Atomic Si and SiO are the major frozen Si-bearing species in the 35
layers closest to the bare surface of the grains, while starting from
$\sim10^5$yr frozen silane and SiO equally dominate. Note that
SiO$_2$ is never an important mantle component, being always at
abundances lower than about $10^{-9}$. We discuss later the
implications for the possible silicate growth.

A first important effect caused by the increase of the CMB temperature
and, consequently, the cloud temperature is that the mantles growth is
slower and the mantles become thinner (Figure
\ref{fig:CMB-Nlayers-time}). At $10^7$ yr, while at 10
K the grain mantles have 122 layers, at 30 K they have 73. Indeed, the increased dust temperature has two
effects: (1) the residence times of H and O atoms, as well as of CO
molecules, are shorter with increasing temperature so that the
hydrogenation and oxidation processes are reduced; (2) when the dust
temperature becomes larger than their sublimation temperatures (linked
to their relative binding energies), CO and CH$_4$
cannot remain frozen on the surfaces. In summary, as the temperature increases, less and less
molecules remain  on the grain mantles and the mantle thickness, consequently, diminishes.

Also the composition dramatically changes with increasing CMB
temperature. This is shown in Figures \ref{fig:CMB-Mantle1},
\ref{fig:CMB-Mantle2} and \ref{fig:CMB-Mantle3}.
We first discuss the major components of the mantle
(Figs. \ref{fig:CMB-Mantle1} and \ref{fig:CMB-Mantle2} ).  Although
water is always the major mantle component, the contribution of
frozen CO$_2$ increases with increasing temperature (because of the
non-residence of CO). Frozen CO and CH$_4$ disappear from the mantle
for temperatures higher than 20--30 K, because of their relatively
low binding energies. Frozen methanol is very abundant in the top 
mantle layers at temperatures $\leq$20 K, but at $\geq$30 K very
little methanol is synthesised on the grain mantles.  Finally, frozen
ammonia decreases with increasing temperature, after a peak at 20 K,
because of the balance of H-atoms diffusion/residence time.

The abundances of Si-bearing molecules trapped into the grain mantles
are shown in Fig. \ref{fig:CMB-Mantle3}. In all cases, SiO is the
major reservoir of frozen silicon, followed by silane. Silane, a
frozen species so far not predicted by models, is as abundant as iced
SiO for temperatures $\leq$20 K, and about 4 times less abundant at
higher temperatures.
Frozen atomic Si is present in appreciable quantities only at 10 K,
since its oxidation and hydrogenation become more efficient with
increasing temperature. Finally, SiO$_2$ only traps a very tiny
fraction of frozen silicon.

\begin{figure*}
\includegraphics[width=14cm]{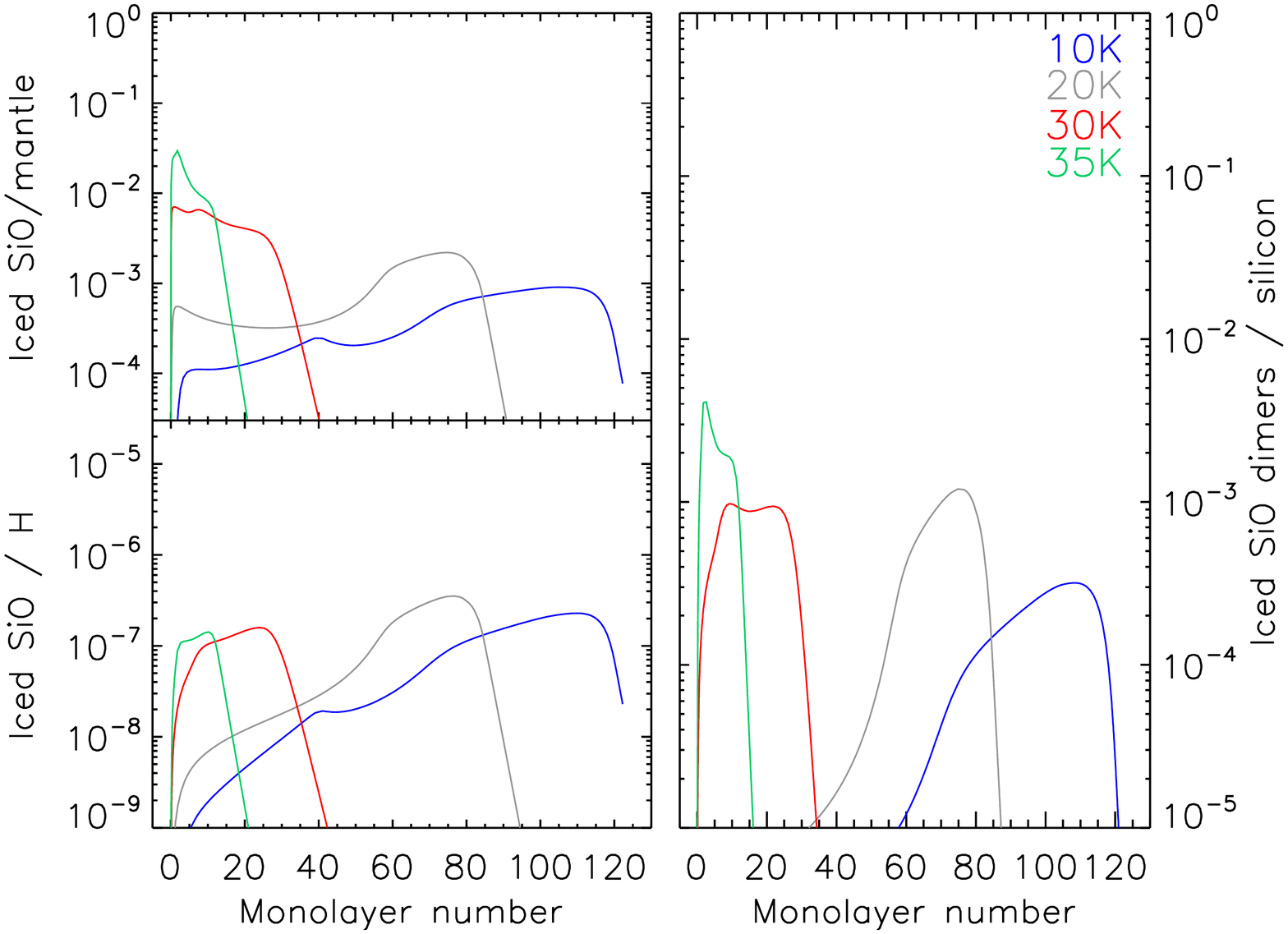}
\caption{{\it Left panels}: Iced SiO over the total frozen species of
  the mantle (top) and iced SiO/H-nuclei (bottom) as a function of the
  monolayers for the four models with different cloud temperature: 10
  K (blue), 20 K (grey), 30 K (red: this is the reference model)
  and 35 K (green). {\it Right panel}: Amount of iced SiO dimers
  normalised to the elemental abundance of silicon in the gas-phase as
  a function of the monolayer, for the four models as above.}
\label{fig:CMB-SiO-H2O}
\end{figure*}

In order to answer the question whether silicates can grow in
molecular clouds, we need to focus on SiO, which might form dimers and
clusters (see Introduction). Specifically, we need to quantify the
probability to have one or more adjacent SiO molecules. Figure
\ref{fig:CMB-SiO-H2O} shows the abundance ratio of frozen SiO over the
total number of frozen species on each layer forming the mantle, for
the four cases with different CMB temperatures. We consider this
ratio as the probability of two SiO molecules to be adjacent: it is
larger than $10^{-2}$ only in the first few layers at a cloud
temperature of 35 K. To quantify how many SiO dimers are statistically
present in the mantle, we first plot the iced SiO abundance with
respect to the total number of H-nuclei in each layer (Fig. \ref{fig:CMB-SiO-H2O}) and
then, multiplying it by the probability that two SiO molecules are
adjacent: this gives the amount of iced SiO dimers in the mantle in
each layer. Fig. \ref{fig:CMB-SiO-H2O} shows this value normalised to
the elemental silicon in the gas-phase. Clearly, the amount of
possible iced SiO dimers is very small, always lower than 0.04 the
gaseous available silicon. The most favourable case is for a cloud at
35 K and in the first $\sim$10 layers of the the mantle. At lower
temperatures, no more than $\sim$0.1\% of elemental silicon is frozen
into SiO dimers. In addition, these dimers are more distant from the
bare silicate surfaces with diminishing cloud temperature.

\subsection{The effect of cosmic ray ionisation rate}\label{sec:effect-zeta}

Theoretical sensitivity studies of the ISM gas phase evolution have
shown that several molecular species are very sensitive to the CR
ionisation rate $\zeta_{CR}$ (e.g. Bayet et al. 2011; Meijerink et al.
2011; Ceccarelli et al. 2011; Bisbas et al. 2015). In this section, we
explore the response of the mantle composition to changes in
$\zeta_{CR}$ of our reference model (Table \ref{tab:parameters}). We
select a range of $\zeta_{CR}$ intended to cover values likely
appropriate for external galaxies. As CR are prevalently accelerated
by SN explosions (e.g. Morlino 2017), the CR ionisation rate is
related to the formation rate of massive stars, and so it is large in
active and early galaxies. Therefore, we varied $\zeta_{CR}$ from the
conservative value for our Milky Way ($1\times 10^{-17}$ s$^{-1}$) up to 1000
times larger ($1\times 10^{-14}$ s$^{-1}$), a value that has been invoked for
the ULIRG Arp220 (Bayet et al. 2011).

\begin{figure*}
\includegraphics[width=14cm]{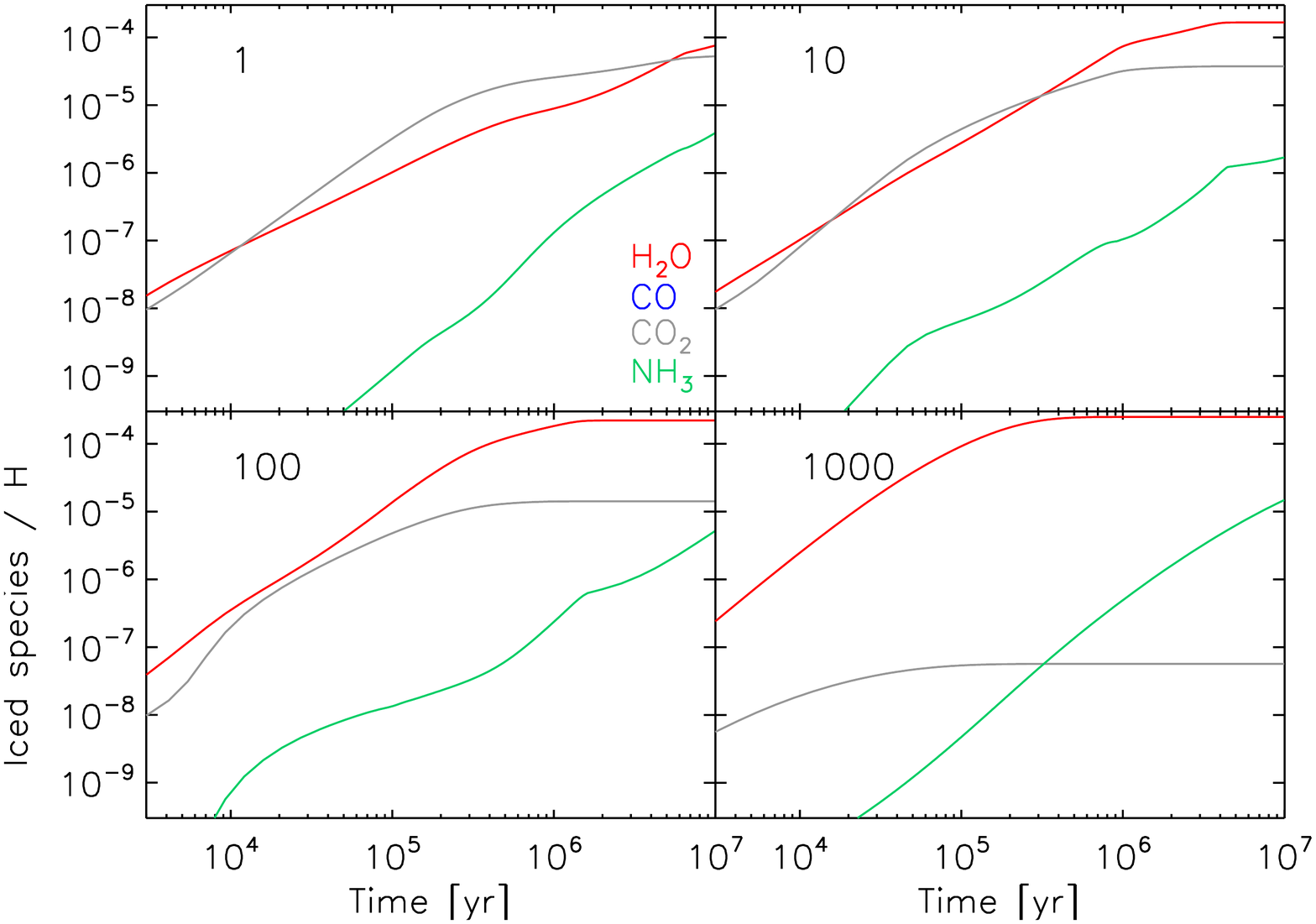}
\caption{As Fig. \ref{fig:CMB-Mantle1}. The simulations refer to
  cases where the CR ionisation rate $\zeta_{CR}$ is varied with
  respect to the value $1\times10^{-17}$ s$^{-1}$: 1 ({\it upper
    left panel}), 10 ({\it upper right panel}), 100 ({\it bottom
    left panel}) and 1000 ({\it bottom right panel}) times larger
  values. The other parameters are those of the reference model (Table
  \ref{tab:parameters}).}
\label{fig:zeta-Mantle1}
\end{figure*}

\begin{figure*}
\includegraphics[width=14cm]{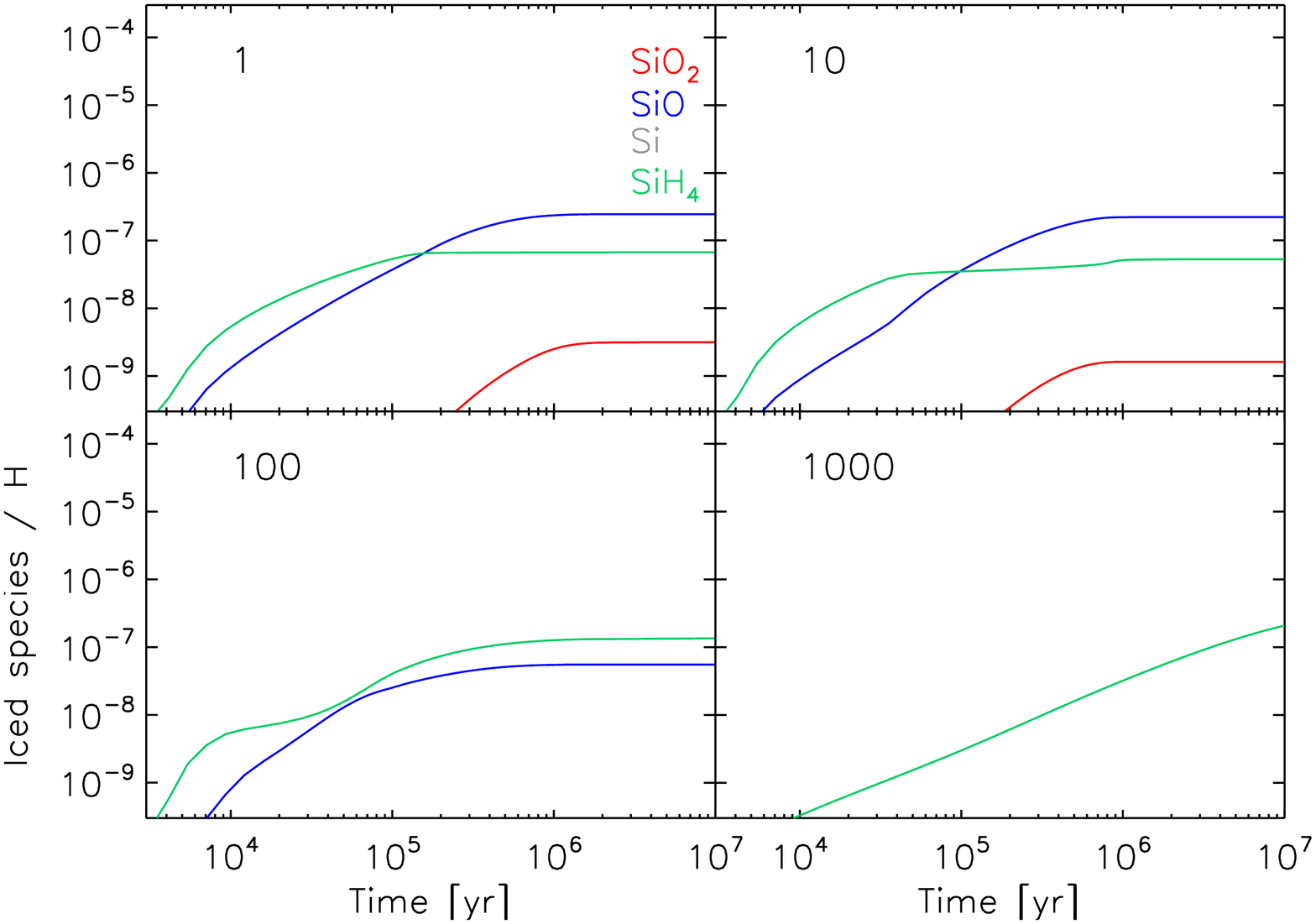}
\caption{As Fig. \ref{fig:CMB-Mantle3}. The simulations refer to
  cases where the CR ionisation rate $\zeta_{CR}$ is varied with
  respect to the value $1\times10^{-17}$ s$^{-1}$: 1 ({\it upper
    left panel}), 10 ({\it upper right panel}), 100 ({\it bottom
    left panel}) and 1000 ({\it bottom right panel}) times larger
  values. The other parameters are those of the reference model (Table
  \ref{tab:parameters}).}
\label{fig:zeta-Mantle3}
\end{figure*}

\begin{figure*}
\includegraphics[width=14cm]{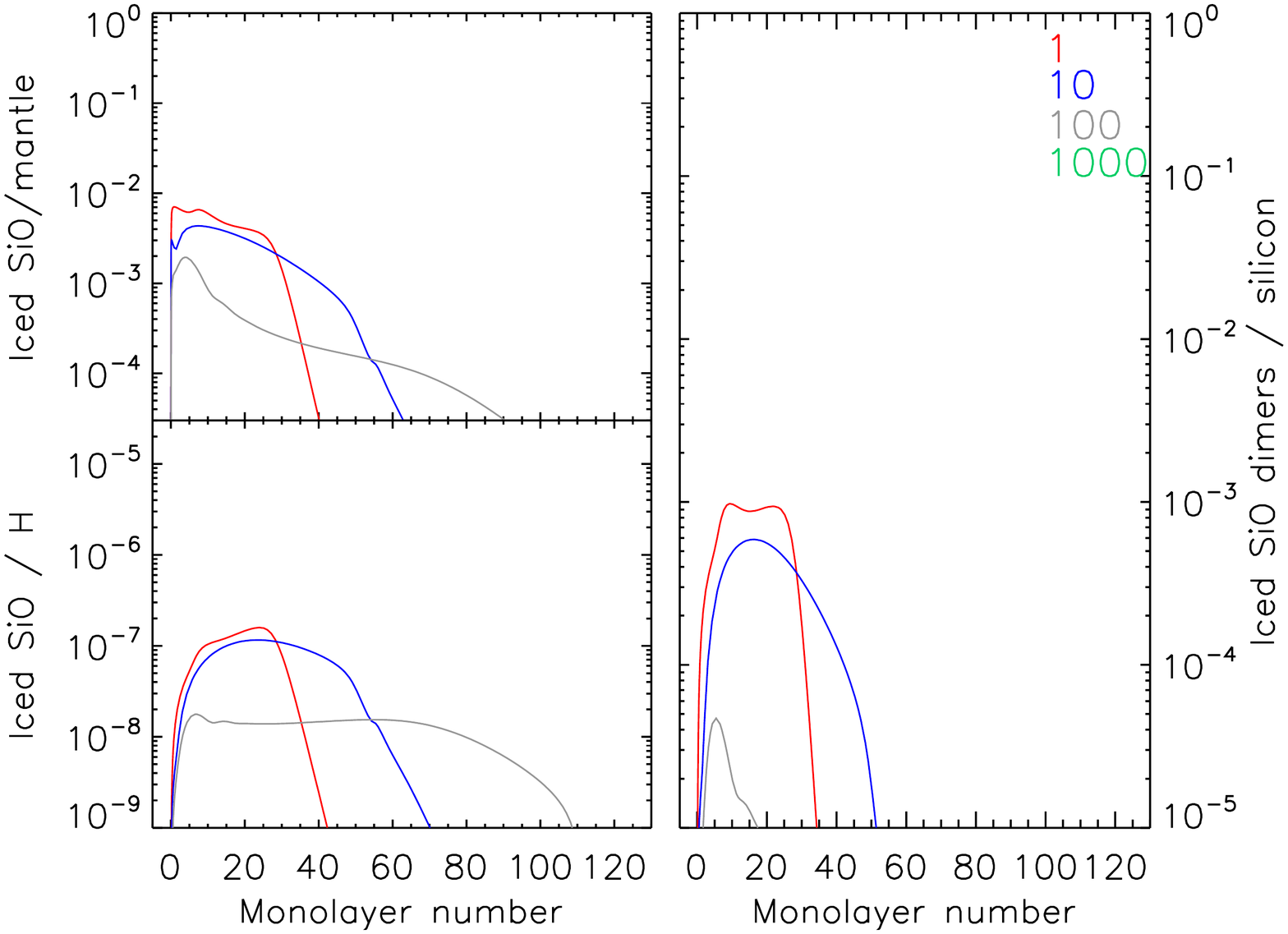}
\caption{As Fig. \ref{fig:CMB-SiO-H2O} for the
  four models with different CR ionisation rate $\zeta_{CR}$: 1 (red),
  10 (blue), 100 (grey) and 1000 (green) times
  10$^{-17}$ s$^{-1}$. }
\label{fig:zeta-SiO-H2O}
\end{figure*}

Figures \ref{fig:zeta-Mantle1} to \ref{fig:zeta-SiO-H2O} show the
results of these simulations.
First, the number of layers increases from 73 (reference model: see
above) to 100, 111 and 112 for $\zeta_{CR}$ equal to 10, 100 and 1000
times the standard value of 10$^{-17}$ s$^{-1}$.
Indeed, increasing $\zeta_{CR}$ leads to an increase in the water ice
abundance as the abundance of atomic O increases because of
destruction of CO in the gas by the CR. While frozen water increases,
CO$_2$ diminishes with increasing cloud temperature. At
$\zeta \geq 10^{-16}$ s$^{-1}$ water dominates over CO$_2$ ice and the
iced CO$_2$ abundance gradually drops from $\sim10^{-4}$ to
$\sim10^{-7}$ at large $\zeta$
(Fig. \ref{fig:zeta-Mantle1}). Similarly, methanol and formaldehyde
(not shown), the other mantles components of the reference model
(Fig. \ref{fig:CMB-Mantle2}), disappear from the mantles with
increasing $\zeta$ as CO is destroyed in the gas by the CR.

Increasing the CR ionisation rate causes the disappearance of SiO$_2$
from the mantles at $\zeta \geq 10^{-15}$ s$^{-1}$ and that of SiO at
$\zeta \geq 10^{-14}$ s$^{-1}$. Starting from $\zeta \sim 10^{-15}$
s$^{-1}$, silane becomes the major reservoir of iced silicon
(Fig. \ref{fig:CMB-Mantle3}).
The ratio of frozen SiO over the total frozen species diminishes with
increasing $\zeta_{CR}$ (Fig. \ref{fig:zeta-SiO-H2O}), leading to
frozen SiO dimers over elemental silicon abundance of less than
$\sim10^{-3}$ in the best case, at low $\zeta_{CR}$. At
$\zeta_{CR}\geq 10^{-15}$ s$^{-1}$ the abundance of SiO dimers over
silicon abundance crashes to $\leq 5\times 10^{-5}$.

We finally note that, in our simulations, we did not include the
possibility that CR sputter the ice mantles, which might
reduce the mantle size, nor that they could induce reactions among
frozen species (e.g. Palumbo \& Strazzulla 1993; Dartois et al. 2015).

\subsection{The effect of metallicity}\label{sec:effect-met}

Metallicity is likely to play a major role in the composition of the
grain mantles. In particular, it may be substantially reduced from the
values of the Milky Way in early, high redshift galaxies. Therefore,
we have explored the response of the icy mantle composition to changes
in metallicity up to 0.01 of the Solar value. We note that for this
study we assume that the dust-to-gas ratio scales with metallicity,
namely we scaled the abundance of all elements, in the gas and in the
dust, by the same amount.

Finally, in order to further reflect the ISM at high redshift, we run
a model where the initial elemental abundances are taken from
theoretical predictions of the yield from the first generation of
massive stars. This is the model $Cosmo$ of Table \ref{tab:elements}.
These yields will depend on the mass of the supernova progenitor and
several models lead to critically different values for such abundances
(e.g. Chieffi \& Limongi 2002; Umeda \& Nomoto 2002; Heger \& Woosley
2002). For this study we choose a 80 M$_{\sun}$ progenitor and use the
yields calculated by the Chieffi \& Limongi (2002) models. As the
yield from supernovae will be diluted by mixing with the pristine
surrounding gas, we normalize the calculated yields to the carbon
abundance of our 0.1 metallicity model and we assume that half of the
carbon and half of the silicon are locked in the grain cores.
Assuming that the bulk of the dust is due to carbonaceous grains (see
below), we adopted a dust-to-gas ratio equal to 1/300, namely 1/3 the
standard one. The initial elemental abundances for this model run are
also listed in the last column of Table \ref{tab:elements}.

\begin{figure*}
\includegraphics[width=14cm]{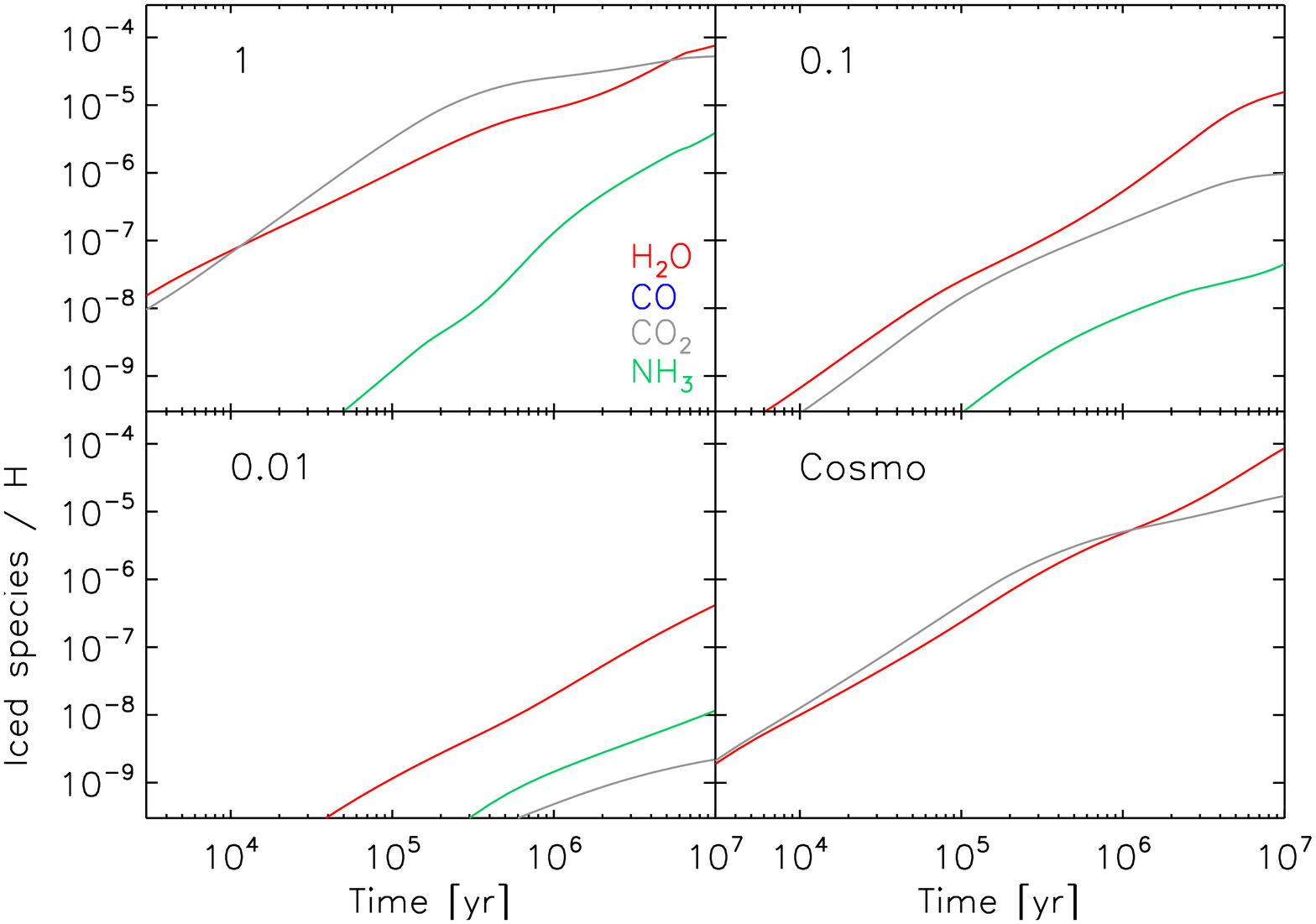}
\caption{As Fig. \ref{fig:CMB-Mantle1}. The simulations refer to cases
  where the metallicity varied with respect to the Solar value: 1
  ({\it upper left panel}), 0.1 ({\it upper right panel}), 0.01 ({\it
    bottom left panel}) times larger values, and from cosmic
  simulations (see text) ({\it bottom right panel}). The other
  parameters are those of the reference model (Table
  \ref{tab:parameters}).}
\label{fig:met-Mantle1}
\end{figure*}

The number of layers slightly increases, from 73 to 82, when the
metallicity decreases by a factor 10, but then it drops to 24 layers
for a metallicity lower than a factor of 100 with respect to the Solar
one. This is mainly due to the behaviour of the atomic oxygen. With
metallicity 0.1 solar, atomic O abundance is always comparable or less than
atomic H, so that water molecules are more easily formed on the surface. However, when the
metallicity drops to 0.01, there are too few oxygen atoms. 
The composition of the mantles is shown Figure
\ref{fig:met-Mantle1}. Water is always the major component, followed
by iced CO$_2$ and ammonia.
The $Cosmo$ simulation gives the most different results, which reflect
the adopted elementary abundances (Table \ref{tab:elements}). For
example, no ammonia is present because of the very low assumed
nitrogen elementary abundance.

\begin{figure*}
\includegraphics[width=14cm]{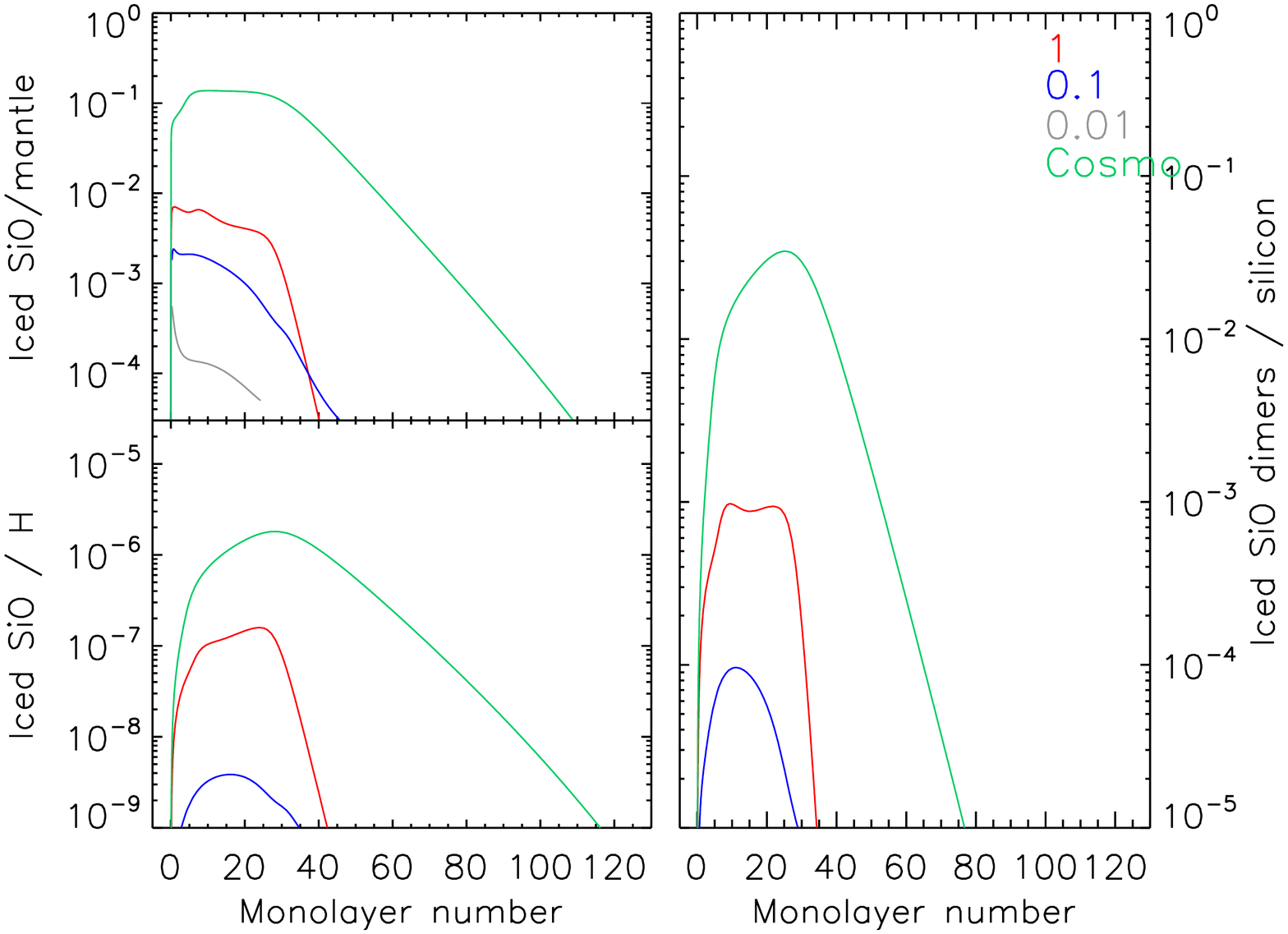}
\caption{As Fig. \ref{fig:CMB-SiO-H2O} for the four models with
  different metallicity: 1 (red), 0.1 (blue), 0.01 (grey) times
  the Solar value, and from cosmic simulations (see text) (green).}
\label{fig:met-SiO-H2O}
\end{figure*}

In all the simulations with varying metallicity, the
abundance of possible SiO dimers is, as in the previous cases, always
lower than $10^{-3}$, with the exception of the $Cosmo$ simulation,
where the gaseous silicon abundance is half of the elemental one. In
this case, the dimers abundance can reach 0.04 the elemental silicon
abundance between 20 and 40 monolayers from the bare silicates (Figure
\ref{fig:met-SiO-H2O}). Hence the Cosmo simulation is the most favourable to the formation of SiO dimers on grains. 

\subsection{The effect of silicon abundance}\label{sec:effect-Si}

One parameter that is certainly crucial in increasing the SiO and
SiO$_2$ abundance in the mantle and, more particularly, in the first
layer(s), is the amount of available gaseous silicon. We, therefore,
chose to vary the initial elemental abundances of silicon in the gas
phase, $Si_{gas}$, from 1/100 to 90\% of the total Si abundance,
$Si_{sol}$, assumed to be the solar one ($3.2\times10^{-5}$).
If the grain abundance were dominated by the silicates, we should
reduce accordingly the dust-to-grain ratio. However, this is not the
case even though it is still debated what is the mineralogical nature
of the interstellar grains (e.g. Jones et al. 2013). We, therefore, 
prefer to keep the same dust-to-gas ratio (1/100), with the goal to
explore whether any silicate-like (mainly SiO dimers and clusters)
could growth on the grain mantles, regardless of the composition of the
bare refractory grain.

\begin{figure*}
\includegraphics[width=14cm]{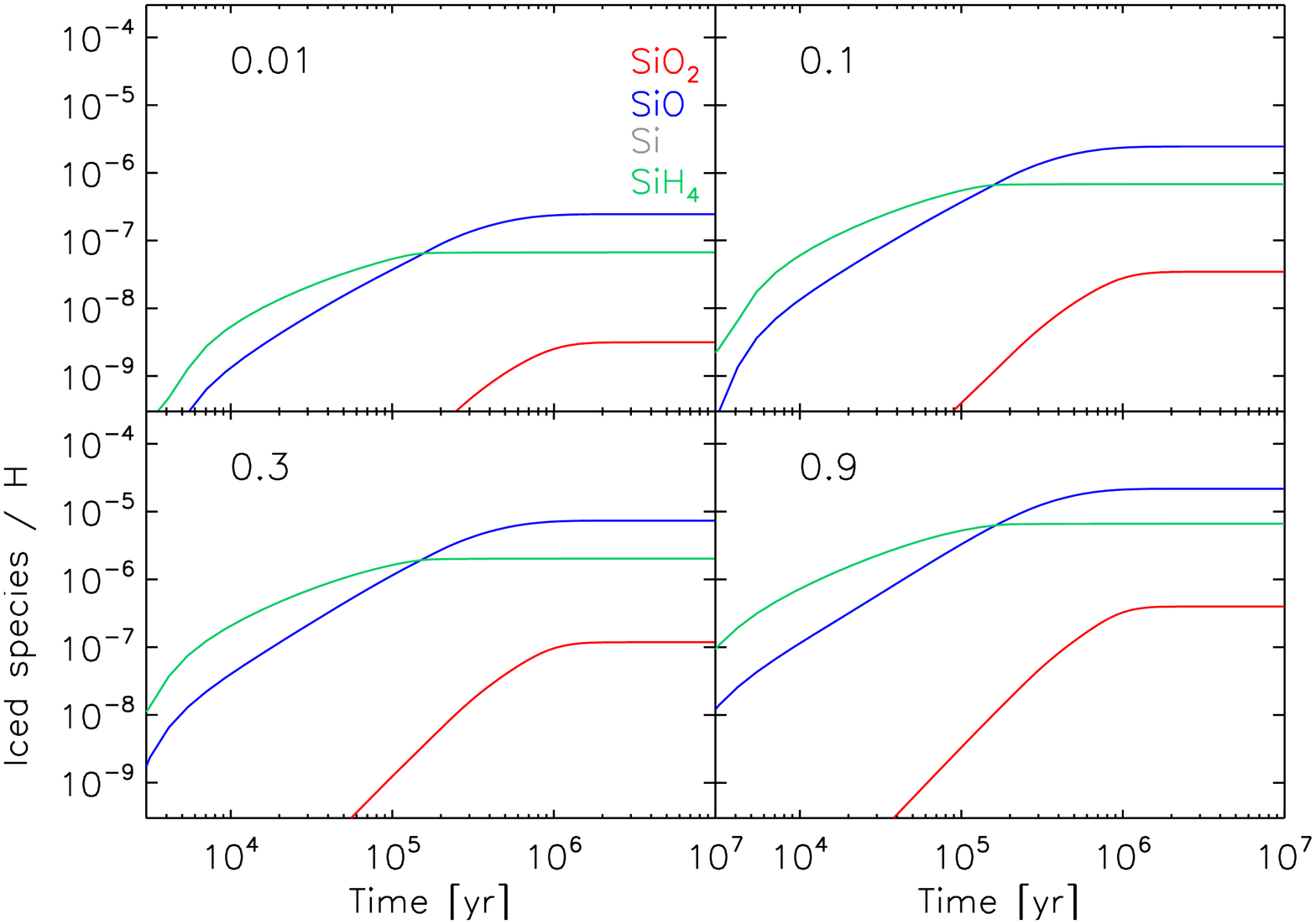}
\caption{As Fig. \ref{fig:CMB-Mantle3}. The simulations refer to
  cases where the gaseous silicon abundance is varied. The
  $Si_{gas}/Si_{sol}$ is 0.01 ({\it upper left panel}), 0.01 ({\it
    upper right panel}), 0.3 ({\it bottom left panel}) and 0.9 ({\it
    bottom right panel}). The other parameters are
  those of the reference model (Table \ref{tab:parameters}).}
\label{fig:Si-Mantle3}
\end{figure*}

\begin{figure*}
\includegraphics[width=14cm]{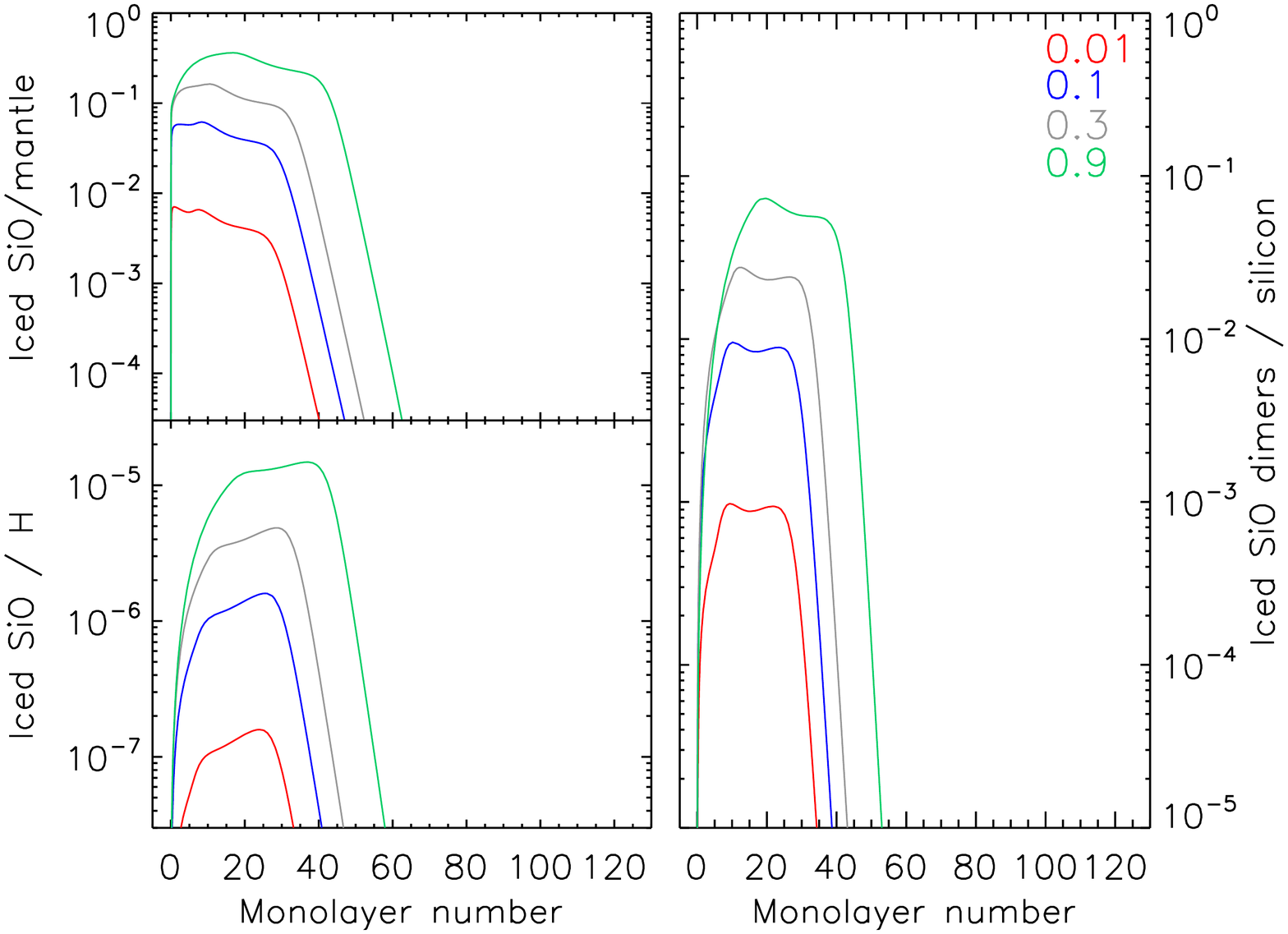}
\caption{As Fig. \ref{fig:CMB-SiO-H2O} for the
  four models with different $Si_{gas}/Si_{sol}$: 0.01 (red),
  0.1 (blue), 0.3 (grey) and 0.9 (green). }
\label{fig:Si-SiO-H2O}
\end{figure*}

The results of these simulations are shown in Figures
\ref{fig:Si-Mantle3} and \ref{fig:Si-SiO-H2O}.
The number of layers in the mantle increases from 73
($Si_{gas}/Si_{sol}$=0.01) to 79 ($Si_{gas}/Si_{sol}$=0.9) because of
gaseous silicon becoming more abundant. Fig. \ref{fig:Si-Mantle3}
shows the increasing amount of SiO, SiH$_4$ and SiO$_2$ with
increasing available gaseous silicon (as expected, the curves of iced
H$_2$O and CO$_2$ do not change). Obviously, increasing the available
elemental silicon helps the production of SiO as well as SiO$_2$ in
the ices.

Nonetheless, even with 90\% of all the silicon in the gas phase the
iced ratio of SiO never exceeds 0.4 times the other species in the
mantle (Fig. \ref{fig:Si-SiO-H2O}). The highest value is found in the
intermediate layers, between $\sim$10 and $\sim$40, where the mantle
is dominated by silane first and later by CO$_2$. 
However, most importantly, the abundance of SiO dimers with respect to
the available elemental silicon never exceeds 0.1 and decreases to
0.01 for $Si_{gas}/Si_{sol}$ equal to 0.1. In other words, even
assuming that 90\% of silicon is in the gas phase, only 10\% of it
will perhaps form SiO dimers. Hence, SiO clusters would have
extremely low abundances.

\section{Discussion}\label{sec:discussion}

\subsection{The composition of grain mantles}\label{sec:comp-grain-mantl}
The simulations reported in the previous Section show that the grain
mantle composition is a function of $z$ as well as other parameters
such as the CR ionisation rate and the metallicity. Table
\ref{tab:tablone} summarises our results.
\begin{table*}
  \begin{tabular}{rr|rrrrrr|rrrr}
    \hline
    \hline
    Model & Number    & H$_2$O & CO & CO$_2$ & NH$_3$ & CH$_4$ & CH$_3$OH & SiO$_2$ & SiO & Si & SiH$_4$ \\
    number & of layers & \multicolumn{6}{|l|}{$\times 10^{-5}$} & \multicolumn{4}{|l|}{$\times 10^{-7}$} \\
    \hline
    \multicolumn{3}{l}{CMB temperature} &&&&&&&&& \\
     1 & 122 & 8.9 & 2.1 & 2.7 & 2.7 & 1.3 & 3.6 & 0.01 & 1.3 & 0.2 & 1.1 \\
     2 & 117 & 8.9 & 1.7 & 6.2 & 6.4 & 1.3 & 2.8 & 0.04 & 1.8 & 0.00 & 1.2 \\
     3 &   73 & 7.6 & 0.00 & 5.3 & 0.4 & 0.00 & 0.00 & 0.03 & 2.5 & 0.00 & 0.7 \\
     4 &   26 & 2.0 & 0.00 & 0.9 & 0.2 &0.00 & 0.00 & 0.1 & 0.3 & 0.00 & 0.6 \\
    \hline
    \multicolumn{4}{l}{Cosmic-ray ionization rate} &&&&&&&& \\
     5 &   73 & 7.6 & 0.00 & 5.3 & 0.4 & 0.00 & 0.00 & 0.03 & 2.5 & 0.00 & 0.7 \\
     6 & 101 & 17 & 0.00 & 3.7 & 0.2 & 0.00 & 0.00 & 0.02 & 2.2 & 0.00 & 0.5 \\
     7 & 111 & 22 & 0.00 & 1.4 & 0.5 & 0.00 & 0.00 & 0.00 & 0.6 & 0.00 & 1.3 \\
     8 & 113 & 25 & 0.00 & 0.00 & 1.5 & 0.00 & 0.00 & 0.00 & 0.00 & 0.00 & 2.1 \\
    \hline
    \multicolumn{3}{l}{Metallicity} &&&&&&&&& \\
     9 &  73 & 7.6 & 0.00 & 5.3 & 0.4 & 0.00 & 0.00 & 0.03 & 2.5 & 0.00 & 0.7 \\
   10 &  82 & 1.6 & 0.00 & 0.1 & 0.04 & 0.00 & 0.00 & 0.00 & 0.07 & 0.00 & 0.07\\
   11 &  24 & 0.04 & 0.00 & 0.00 & 0.00 & 0.00 & 0.00 & 0.00 & 0.00 & 0.00 & 0.00 \\
   12 &141 & 8.6 & 0.00 & 1.7 & 0.00 & 0.00 & 0.00 & 0.9 & 27.8 & 0.00 & 2.7 \\
    \hline
    \multicolumn{3}{l}{Silicon abundance} &&&&&&&&& \\
   13 & 73 & 7.6 & 0.00 & 5.3 & 0.4 & 0.00 & 0.00 & 0.03 & 2.5 & 0.00 & 0.7 \\
   14 & 73 & 7.5 & 0.00 & 5.2 & 0.4 & 0.00 & 0.00 & 0.3   & 25  & 0.00 & 6.8 \\
   15 & 75 & 7.2 & 0.00 & 5.2 & 0.4 & 0.00 & 0.00 & 1.2   & 74  & 0.00 & 20  \\
   16 & 79 & 6.5 & 0.00 & 5.0 & 0.4 & 0.00 & 0.00 & 4.0   & 2.2 & 0.00 & 66 \\
    \hline
    \hline
  \end{tabular}
  \caption{Summary of the main results of the simulations. The
    abundances of frozen H$_2$O, CO, CO$_2$, NH$_3$, CH$_4$ and
    CH$_3$OH are in units of $10^{-5}$ with respect of H nuclei,
    whereas the abundances of SiO$_2$, SiO, Si and SiH$_4$ are in
    $10^{-7}$ units.}\label{tab:tablone}
\end{table*}

Water and carbon dioxide are always the most abundant iced species
whereas methane, carbon monoxide, methanol and formaldehyde are
extremely sensitive to the cloud temperature, namely the redshift of
the galaxy to which the cloud belongs to. Ammonia represents an
intermediate case, sensitive to the cloud temperature, the CR
ionisation rate and (not surprising) metallicity. Iced silane and SiO
are always the major reservoirs of silicon in the grain mantles,
whereas atomic silicon and SiO$_2$ contain a very small fraction, at
most a few percent, of it.

Conversely, the grain iced mantles in distant galaxies might provide
crucial information on these different parameters. Perhaps, the most
difficult to evaluate by means of observations is the CR
ionisation rate $\zeta_{CR}$. Our simulations show that the relative
abundance of iced water and carbon dioxide might provide this
information, with the iced H$_2$O/CO$_2$ decreasing with increasing
$\zeta_{CR}$. The James Webb Space Telescope (JWST) with its great
sensitivity in the NIR, where these two species have ice features, should
be able to constrain $\zeta_{CR}$ across the galaxies where the
observations are feasible, once the other parameters are known via
different observations.

Very few iced H$_2$O, CO and CO$_2$ have been so far detected in
external galaxies. Ices have been observed in the Large and Small
Magellanic Clouds, LMC and SMC, where, on average, the gas and dust
are warmer than in the Galaxy, with dust temperatures around 20-40 K
(e.g. van Loon et al. 2010 and references therein) and are metal
poorer (e.g. Pagel 2003 and references therein). In the LMC, iced
CO$_2$/H$_2$O is about twice larger than in the Galaxy (e.g.
Shimonishi et al. 2010; Seale et al. 2011), in (rough) agreement with
our predictions (Table \ref{tab:tablone}). In SMC, the iced
CO$_2$/H$_2$O is slightly smaller than in the Galaxy (Oliveira et al.
2013), again in rough agreement with our predictions when also the
smaller metallicity is considered. Recent AKARI observations have
revealed the presence of H$_2$O and CO$_2$ ices in several nearby
star-forming galaxies (Yamagishi et al. 2013, 2015). They find that
indeed the ratio of these two species is very region dependent and
that it is not easily correlated to a single parameter. In particular
they find that while H$_2$O ice is detected in all their sample,
CO$_2$ ice is only detected in less than half of the sample and its
abundance seems to be correlated with the average dust temperature of
the galaxy, in agreement with our predictions.

Clearly, dedicated modelling is needed to
better refine the agreement between the observed and predicted iced
mantle abundances, which is not the scope of this article.

One indirect consequence of our simulations is the prediction that
interstellar complex organic molecules (iCOMs: Ceccarelli et al. 2017)
may be difficult to be synthesised in high $z$ galaxies. In fact,
methanol is believed to be a crucial species for the synthesis of
iCOMs (e.g. Garrod \& Herbst 2006; Balucani et al. 2015; Taquet et al.
2016). In the cold molecular clouds of our Galaxy, methanol is mostly
synthesised on the grain surfaces via the hydrogenation of CO (e.g.
Rimola et al. 2014). However, for $z$ larger than about 4, CO cannot
remain frozen on the grain surfaces so that no methanol can be
synthesised and, consequently, many of the detected galactic iCOMs
might have much smaller abundances. At present, a few iCOMs have been
detected in extragalactic sources (e.g. {\it
  http://www.astro.uni-koeln.de/cdms/molecules}) and the most distant
galaxy where they were detected is at z=0.89 (M\"uller et
al. 2013). It will be interesting to see whether our model predictions
are correct. If they are, planets and comets of high-zeta galaxies
might develop a very different organic chemistry with respect to that
of the Solar System.

\subsection{Can silicate dust grow in high redshift molecular clouds?}\label{sec:can-dust-grow}

It is clear from the results reported in Sec. \ref{sec:results} that
growing silicate dust is not efficient in any of the parameter space
investigated, as the key species, SiO and Si, needed to form
dimers and clusters from which a silicate might form are always a very
minor iced Si-bearing species. Therefore, our basic conclusion is that
{\it only a minor fraction of gaseous silicon can agglomerate into the
  dust icy mantle to possibly growth silicates.}  Even in the best
case, when 90\% of silicon is in the gas phase (Figure
\ref{fig:Si-SiO-H2O}), only 10\% of it would be able to form SiO
dimers. This is just due to the low silicate abundance with respect to
oxygen, carbon and hydrogen, which means that the dust mantles are
dominated by O- and C- bearing species.

So far we have not discussed the possibility that carbonaceous grain
can growth in molecular clouds. There is now consensus that
carbonaceous dust is formed by graphite with different degree of
hydrogenation (e.g. Jones et al. 2013). The growth of graphite is even
less probable than silicate, as carbon atoms are very rare in the ISM,
and in particular in molecular clouds, due to the fact that carbon
easily goes from the ionised state to the molecular one (CO), and there is just a small
parameter space where it remains atomic (in the so-called Photo
Dissociation Regions, PDRs). As a matter of fact, it is believed that
carbonaceous dust is formed in the AGB and SNe, and that in PDRs it is
fragmented and gives rise to the PAHs and the small carbonaceous
hydrogenated grains (Jones et al. 2013, 2017). In this respect,
therefore, even the growth of carbonaceous dust seems improbable.

\section{Conclusions}\label{sec:conclusions}

In this paper we present a theoretical study on the ice mantle
evolution as a function of time and physical parameters
representative of extragalactic environments at different
redshifts. Our aim is two-fold: firstly, we explore how the main ice
composition varies with changes in temperatures, cosmic ray ionization
rates, metallicity and the amount of silicon in the gas; secondly, we
quantify whether grain growth at high redshift can indeed be efficient
enough to account for the presumed missing dust mass.  

Our main conclusions are the following:\\

\noindent
1) The composition of the iced mantles is a strong function of the
  redshift of the galaxy to which the molecular cloud belongs to. The
   parameter that affects this composition most is the temperature of the dust,
  which increases with increasing redshift. While water is always the
  major mantle component, the relative abundances of CO and CO$_2$, the
  major other mantle components in the galactic clouds, strongly
  depend on the
  dust temperature, with frozen CO disappearing at $z\geq 4$.\\

\noindent
2) Methanol is unlikely to be abundant in the grain mantles of
  molecular clouds at redshifts higher than about 4. This also implies
  that complex organic molecules are unlikely to be abundant in
  the star forming regions of those galaxies, with possibly profound
  implications on the organic chemistry of the nascent planetary
  systems.\\

\noindent
3) The CR ionization rate $\zeta_{CR}$ and metallicity are also
  extremely important in the final composition of the grain iced
  mantles. In particular, the frozen H$_2$O/CO$_2$ ratio increases with
  decreasing $\zeta_{CR}$.\\

\noindent
4) Even the most favourable scenario to grain growth, namely
$\sim$90\% of the elemental Si in the gas phase and high metallicity,
is unlikely to lead to a significant grain growth and, consequently,
to any significant contribution toward the presumed missing dust mass
problem.

\section*{Acknowledgements}
We warmly thank Raffaella Schneider and Ilse De Looze for insightful
discussions on the dust growth in Supernovae explosions and, in
general, in the high redshift clouds.  
We acknowledge funding from the European Research Council (ERC) in the
framework of the ERC Advanced Grant Project DOC ``The Dawn of Organic
Chemistry'' GA N. 741002.





\bsp 

\label{lastpage}

\end{document}